\documentclass[10pt]{article} 
\usepackage[english]{babel}
\usepackage{graphicx}
\usepackage{amsmath}
\usepackage{amssymb}
\usepackage{mathtools}
\usepackage{braket}
\usepackage{setspace}
\usepackage[utf8]{inputenc}
\usepackage[makeroom]{cancel}
\usepackage{enumerate}
\usepackage{cite}
\usepackage{float}
\usepackage[symbol]{footmisc}
\usepackage{placeins}
\usepackage[misc]{ifsym}

\usepackage[top=2.5cm, bottom=2.5cm, left=2.5cm, right=2.5cm]{geometry}

\usepackage{color}
\usepackage[colorlinks=true,
            linkcolor=red,
            urlcolor=blue,
            citecolor=lightblue]{hyperref}
\definecolor{darkblue}{rgb}{0.0, 0.0, 0.5}
\definecolor{darkblue}{RGB}{0,0,80}
\definecolor{lightblue}{RGB}{50,120,150}
\definecolor{darkgreen}{RGB}{0,80,0}
\definecolor{darkred}{RGB}{80,0,0}
\definecolor{amber}{rgb}{1.0, 0.75, 0.0}
\definecolor{arsenic}{RGB}{50,30,90}
\definecolor{ao(english)}{rgb}{0.0, 0.5, 0.0}

\usepackage{xparse}
\usepackage{cleveref}

\renewcommand{\thefootnote}{\fnsymbol{footnote}}

\title{\textsc{Plant efficiency: \\
a sensitivity analysis of the capacity factor for fusion power plants with high recirculated power}}

\author{Authors: R. A. Mulder\footnote{Science and Technology of Nuclear Fusion, Eindhoven University of Technology, Eindhoven, The Netherlands.}  \footnote{Corresponding author: \, \Letter  \,  ruwardarthur@gmail.com or ram202@cam.ac.uk.} , Y. G. Melese$^{*}$, N. J. Lopes Cardozo$^{*}$}
\date{}
\begin{document}
\renewcommand{\thefootnote}{\arabic{footnote}}
\maketitle 
%\newdate{date}{06}{06}{2018}

\begin{abstract}
\noindent The plant efficiency of a nuclear fusion power plant  is considered. During nominal operation, the plant efficiency is determined by the thermodynamic efficiency and the recirculated power fraction. However, on average the reactor operates below the nominal power, even when the long shutdown periods for large maintenance are left outside the averaging.  Hence, next to the recirculated power fraction, the capacity factor must be factored in. An expression for the plant efficiency which incorporates both factors is given. It is shown that the combination of high recirculated power fraction and a low capacity factor results in poor plant efficiency. This is due to the fact that in a fusion reactor the recirculated power remains high if it runs at reduced output power. It is argued that, at least for a first generation of power plants, this combination is likely to occur. Worked out example calculations are given for the models of the Power Plant Conceptual Study. Finally, the impact on the competitiveness of fusion on the energy market is discussed. This analysis stresses the importance of the development of plant designs with low recirculated power fraction.
\end{abstract}

\section{Introduction}
Nuclear fusion holds the promise of safe, clean and CO$_2$-free and virtually unlimited energy. Yet, it still has a number of serious technological problems to overcome before it can be deployed as a commercial power source. Many of these will be addressed by the international ITER experiment, presently under construction. After ITER, the EUROfusion roadmap foresees the construction of a DEMO reactor, after which the scene is set for the first generation of commercial fusion power plants \cite{EUROFusion}. An analysis of what such reactors might look like was presented in the comprehensive EFDA Power Plant Conceptual Study (PPCS) \cite{PPCS}. Herein, a range of reactor models A to D is considered, where the models A and B are based on physics and technology close to what is presently available, whereas the more advanced models C and D make significant extrapolations. All models are based on the steady state tokamak concept, which requires a non-inductive drive of the toroidal plasma current. In this paper, we shall use the PPCS as reference.
The early models of these reactors will be characterised by a high recirculated power fraction ($f_{recirc}$), which is the power that is taken from the gross electric output and recirculated back to the operation of the power plant itself, mainly consisting of the power needed to drive the plasma current, to pump the coolant and to run the cryoplant. A high recirculated power is detrimental to the plant efficiency and economy, a point that has been made in the literature, mostly as an incentive for the development of efficient current drive systems \cite{PPCS, Zohm, Pamela, Buttery, Stork}.

In this paper we introduce a new element into the discussion, namely that the effective -- averaged over time -- output power will necessarily be lower than the nominal or 'nameplate' power. 
For most energy sources it is normal to on average run below the nominal power. It is expressed by the capacity factor $\chi$, i.e., the time-averaged power normalised to the nameplate power. As long as the recirculated power is a small fraction of the output power, the capacity factor has little influence on the plant efficiency ($\eta_{\text{plant}}$). For example, intermittent sources such as wind or solar have a low capacity factor, but they have almost no recirculated power.  For a fusion power plant, of which the power consumption remains high when it runs at an output power below its design value, the situation is different. Combining a high recirculated power fraction with an average output below the nominal power operation, %especially the early models of the
 fusion power plants are liable to operate in a regime in which the plant efficiency, and hence the net electricity production, is much lower than presently assumed.  

In this paper we are primarily interested in the first generation commercial power plant, which are most likely to suffer from the combination of high recirculated power and relatively low capacity factor. But we will  keep the discussion as general as possible, using values from PPCS, DEMO and other publications to guide our estimations.

\begin{figure}
\centering
\includegraphics[width=0.6\textwidth]{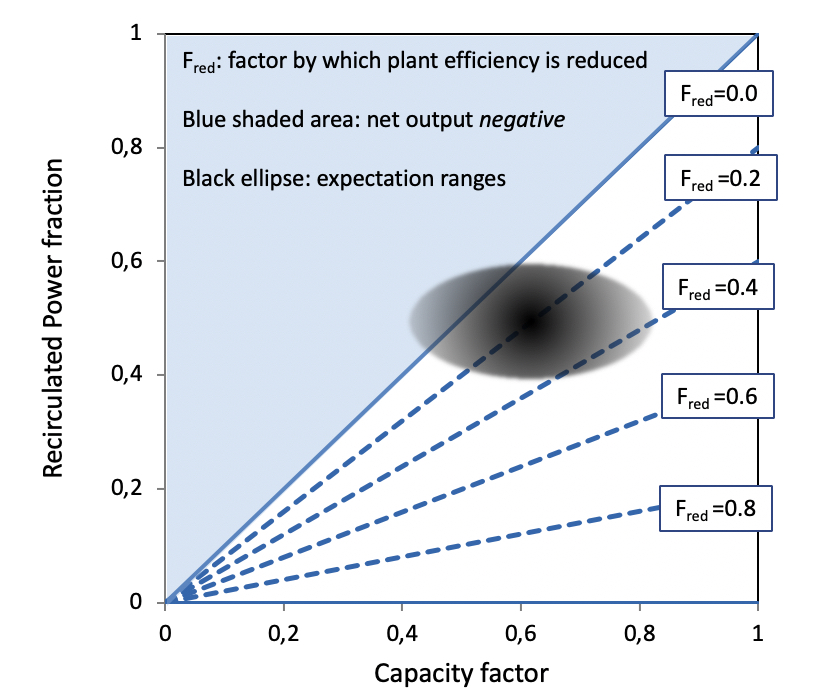}
\caption{\small{The capacity factor and recirculated power fraction determine the reduction of the plant efficiency according to Eq. \eqref{efficiencyreduction}. The shaded area represents the expectation values  of fusion power plants based on the steady state tokamak concept, as estimated in  sections 2 and 3 of this paper. Note how the combination of a high recirculated power and a modest capacity factor results in serious reduction of the overall plant efficiency. In contrast, with a low recirculated power, more typical of conventional power plants or renewable energy technologies, the plant efficiency does not depend on the capacity factor.}} \label{fig:1}
\end{figure}

For a generic power plant with conversion efficiency $\eta_{\text{conv}}$, recirculated power fraction $f_{\text{recirc}}$, and capacity factor $\chi$, the plant efficiency is given by\footnote{It is shown in section \ref{PE} how this formula is constructed from first principles. }
\begin{equation}\label{efficiencyreduction}
\eta_{\text{plant}} =  \eta_{\text{conv}} (1 - f_{\text{recirc}} /\chi),
\end{equation}
where the factor $F_{\text{red}} = (1 - f_{\text{recirc}} /\chi)$ gives the reduction of the plant efficiency compared to the nominal value. Figure \ref{fig:1} shows how $F_{\text{red}}$ depends on $f_{\text{recirc}}$ and $\chi$, showing that as long as the recirculated power fraction is small, the plant efficiency is not affected much by the capacity factor. Conversely, the capacity factor is of great importance for a plant with a high recirculated power fraction.\footnote{The capacity factor and the recirculating power are approximated as being independent: the tokamak runs at the same plasma current and magnetic field independent of its output power. However, the bootstrap current and current drive efficiency do depend in non-trivial ways on the temperature and density of the burn equilibrium and the (less important) pumping power on the output power, but for the analysis in this paper these effects can be neglected.\label{firstapprox}} A plant consuming half of its nominal output power to run itself, does not produce any net electricity if it runs at half of its nominal capacity. But also for more favourable values of $f_{\text{recirc}}$ and $\chi$, the reduction of the plant efficiency can be substantial. The graph also shows the expectation values for both quantities which we will estimate in this paper.

%Fusion power plants might be the first power source for which the recirculated power fraction is so high that the capacity factor becomes important for the plant efficiency. 

Since  steady state fusion reactors of the tokamak family have relatively high $f_{\text{recirc}}$, it is important to evaluate how this works out in combination with a realistic estimate of $\chi$. To evaluate these parameters, we will make a breakdown of the operation into periods of operation (which can be at less than nominal power, but require full recirculated power), short periods of maintenance (during which there is no power production yet still some power is needed to run the plant) and the long shutdown periods when the plant is switched off, e.g., for the replacement of the blanket. With the results of this analysis we can estimate expected values of $\chi$ and $f_{\text{recirc}}$, and hence the power plant efficiency. Finally, we discuss the consequences for the competitiveness of the fusion power plant on the energy market. 

%It is important for the analysis to note that the fusion power plants differ from other power plants in two ways. First, they are expected to have regular long shutdown periods for the replacement of  reactor parts (the blanket and the divertor). While these periods do contribute to the overall availability factor, and hence the economics of the plant, they do not influence the plant efficiency as during the shutdown the plant does not consume energy. Therefore the analysis in this paper is restricted to the periods between the scheduled long maintenance shutdowns. Second, the recirculated power, and in particular the power needed to drive the current in the reactor, are not a function of the output power; the recirculated power is to good approximation the same whether the plant runs at full or reduced output power.  This latter characteristic is particularly important for the effect of capacity factor on the plant efficiency.

\FloatBarrier
\section{Availability, Maintenance and Capacity Factor}\label{AvCap}

\subsection{Definitions}
\begin{figure}[h]
\centering
\includegraphics[width=0.8\textwidth]{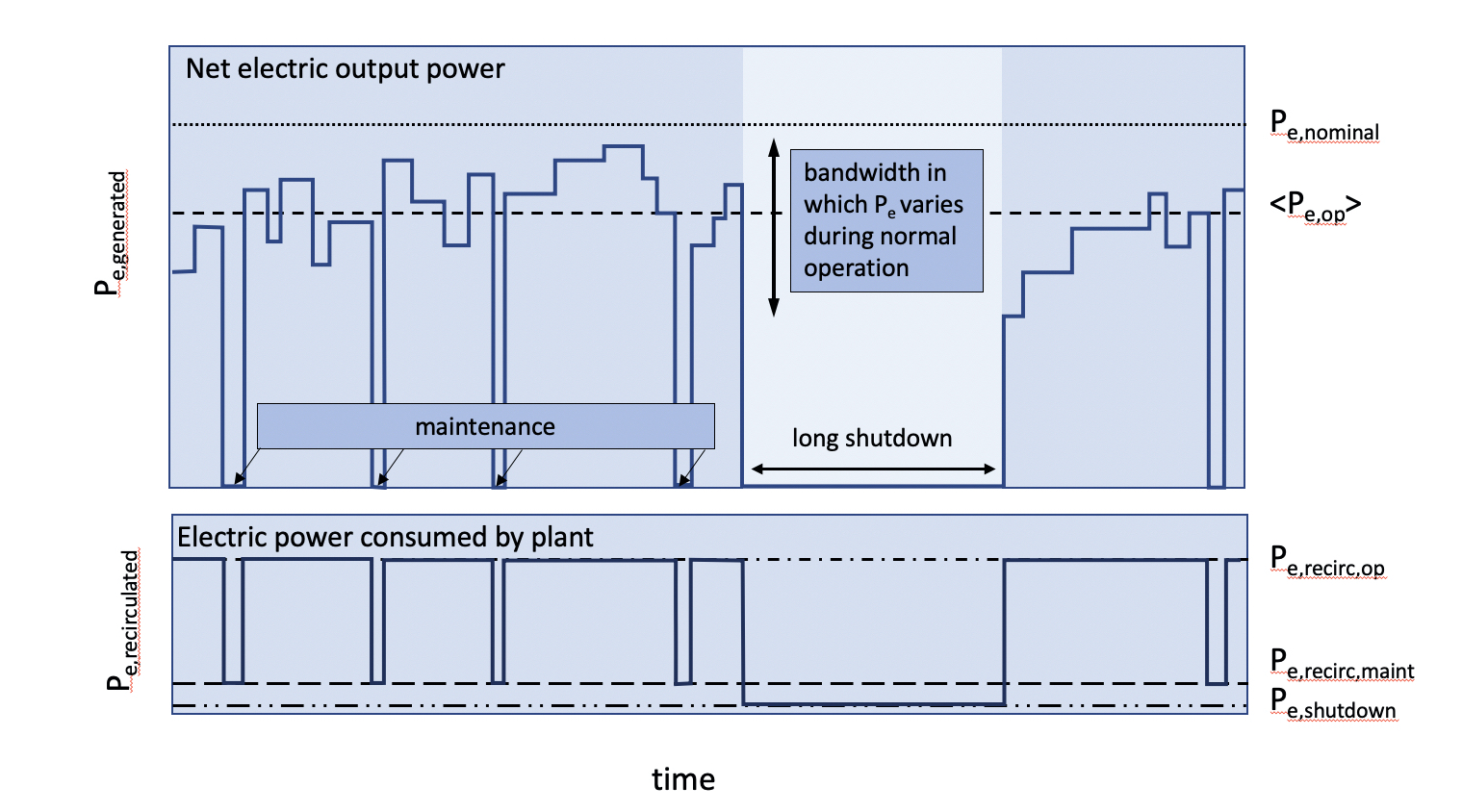}
\caption{\small{Schematic of the operation of the fusion power plant, distinguishing periods of long shutdown (e.g., for periodic blanket replacement), short shutdown (for maintenance) and power production. For several reasons, the output power varies in the latter periods, resulting in an average output power lower than the nominal value. In this paper the capacity factor is evaluated as the average output power during periods of availability, normalised to the nominal output. The effective recirculated power is the average over the same period. For completeness the figure also indicates that also during long shutdown periods there will be some power consumption.}} \label{fig:2}
\end{figure}

Figure \ref{fig:2} summarises the definitions  we will use in the following. These are in agreement with those commonly used in the literature, with an extra specification: a fusion power plant is expected to have regular periods of planned shutdown for the replacement of the divertor and/or blanket modules. The fraction of time the reactor is ready to operate is captured by the `Availability'. The PPCS assumes this Availability to be in the range 0.7--0.8. In this paper we  apply the analysis of the plant efficiency \textit{only to the periods when the plant is available}. %But the Availability does feature in the discussion of the plant economics.\
Yet, in periods of availability the reactor will not run at nominal power at all times, as indicated in the schematic. This is captured by the capacity factor $\chi$ -- ranging from 0 to 1 -- which is defined 
%as the actual output energy over a period of time divided by the energy that would have been produced at nominal capacity %(also referred to as the `rated load') 
%over the same period of time: 
as the energy generated over a period of time, normalised to the energy that would have been generated had the plant run at nominal capacity all the time. This definition is commonly based on the net electric output. However, since in the analysis below the net electric output will depend on the fraction of recirculated power, we choose to define $\chi$ on the basis of the thermal power, so as to keep the two variables independent. 
%For plants with small recirculated power there is little difference. 
Hence,
\begin{equation}\label{CapFac}
\chi =  \Braket{P_{\text{th}}} /P_{\text{th,nom}},
\end{equation}
\noindent where $P_{\text{th}}$ denotes the thermal power of the plant, $P_{\text{th,nom}}$ the nominal thermal power and the brackets denote averaging over an operation period (i.e., when the plant is available). In the next section, we discuss reasons why a reactor would be operated at lower-than-nominal power. 

%\begin{equation}\label{CapFac}
%\chi =  \Braket{P_{\text{e,net}}} /P_{\text{e,nominal}},
%\end{equation}

%\noindent where $P_{\text{e,net}}$ denotes the net electric output power of the plant, $P_{\text{e,nominal}}$ the nominal or nameplate electric output power and the brackets averaging over an operation period.\

During periods of availability there will still be short periods in which the plant is in shutdown, e.g., for maintenance or small repairs. We denote the fraction of the time the plant is not productive due to -- planned or unplanned -- maintenance by $\mu$, stressing that this only refers to short shutdowns in operational periods, and  is therefore to be distinguished from the Availability.

%There is a reference that estimates this.

During such short shutdowns the power consumption of the plant is reduced, since there is no power needed to drive the plasma current, but not zero, since the cryoplant and possibly the cooling pumps and other auxiliary systems will continue to run. This will be absorbed in the averaged recirculated power fraction.

\subsection{Causes for the capacity factor to be lower than one}
While the fusion plant is designed to operate at nominal power as much as possible, it is unavoidable that in practice the time-averaged power output, without counting the periods when the plant is not available, is less than that. Whereas the availability is well discussed in the literature \cite[cf.]{RAMI}, the fact that fusion power plants will on average run below their nominal power is not. Below we review several factors -- some intentional and some unavoidable -- that may reduce the capacity factor of a fusion plant:
\begin{itemize}
\item \textbf{Load-following} (intentional). If the network demands that the power plant has some degree of flexibility to follow demand, that brings down the average power. Load following always lowers the average output; variations above the nominal power are not allowed for machine safety reasons. Fusion power plants could, technically, vary their output power \cite{PPCS}, although this is not trivial, as discussed in Ref.  \cite{WardKemp}. Yet, some capability of demand following could be desirable, if only to stabilise the network and reduce overall system costs \cite{Nicholas}. As a reference we recall that sources such as coal, natural gas and geothermal energy have capacity factors around 50$\%$, mostly because these are used to match demand fluctuations \cite{EIA2018}. Storage could mitigate the effect of fluctuating demand, but may be costly and inevitably leads to losses due to conversion and leakage.

\item \textbf{Operating conditions }(intentional). The nominal output power is realised when the fusion reactor is operated close to the operational limits, especially the density and $\beta$-limits. For instance, the PPCS reactor models have ambitious values for the working point, compared to presently known operational limits. Any deviation from this lowers the output power below the nominal level. Yet the power needed to run the reactor, in particular to drive the current and to cool the magnets, remains the same. One may think of the period of starting up at the beginning of each operational period, after each shut-down period, when the plant will gradually be brought up to its full potential.

\item \textbf{Small maintenance} (planned and unplanned). Periods during which there is no energy generation, but some of the energy-consuming subsystems -- most notably the cryoplant and possibly the coolant pumps -- will continue to operate \cite{Henry}. This is captured by the maintenance parameter $\mu$.

\item \textbf{Deviation from ideal operating conditions} (not intentional). It may not be possible, in practice, to realise the design values of all operational parameters at all times. For instance, a deviation of the ideal fuel mix will reduce the fusion power. Although the time and volume-\textit{averaged} fuel mix can be tuned by a control system, it may not be possible to have the best mix everywhere at all times. In particular, the means to control the fuel mix when fuelling with relatively large pellets are limited. Other factors that could influence the operations include the impurity content, which may evolve during an operational period due to wear; and subtle effects in the transport of the reaction product helium \cite{Jakobs}. Another factor of relevance is the degradation of the superconducting magnets under cycling. Since the fusion power scales as the magnetic field to the fourth power, even a few percent reduction of the maximum field will subtract significantly from the achievable output power. 
\end{itemize}

\subsection{Estimation of the expectation value of the capacity factor}
Having distinguished these factors, we may use them to estimate a reasonable and realistic expectation value for the capacity factor of a fusion power plant. To get an impression, we put forward that for each of the different causes a 10$\%$ reduction of output power would constitute a success. Yet, this would make the plant run at about 65$\%$ of the nominal power. Especially for the early generations of fusion power plants this may not be a pessimistic estimate. To place this in perspective, we note the capacity factor of  fission plants was below 50$\%$ in 1970, some 20 years after their introduction, and has gradually improved over the 40 years that followed to reach the present 80--90$\%$ \cite{NEA}. In recent studies for the EuroFUSION DEMO design, an expectation value of the capacity factor of only 0.2--0.3 is assumed \cite{Coleman, Federici}. While DEMO precedes the first generation of power plants, bringing the capacity factor up from 0.2--0.3 in the demonstrator to a much higher value for the first generation of commercial power plants will be challenge. Thus, 50$\%$  is not an unreasonable expectation value  for the capacity factor of early commercial fusion power plants. 

 In conclusion, there are several reasons why a fusion reactor, on average, will run at a power significantly lower than its nominal power. Because of the large uncertainty in estimating the capacity factor, we introduce the capacity factor ${\chi}$ as a variable in our analysis, so that we can evaluate its influence on the overall plant efficiency.

\FloatBarrier
\section{Effective recirculated power fraction} \label{RPF}
\subsection{Definitions and breakdown}
With reference to the PPCS, we consider the case of a continuously running tokamak fusion reactor which produces a nominal thermal power $P_{\text{th,nom}}$, resulting in a nominal gross electrical power output $P_{\text{e,gross,nom}}$ given by
\begin{equation}\label{GrossP}
P_{\text{e,gross,nom}}= \eta_{\text{conv}}  P_{\text{th,nom}}.
\end{equation}
\noindent The total thermal power is generated by the fusion reaction $P_{\text{fusion}}$ in the reactor core and the additional power $P_{\text{blanket}}$,  generated in the breeding blanket, i.e.,
\begin{equation}\label{ThermalP}
P_{\text{th}} = P_{\text{fusion}}  + P_{\text{blanket}}.
\end{equation}
\noindent The recirculated power $P_{\text{e,recirc}}$ is the electric power that the plant itself uses to sustain its operation.  With reference to Figure \ref{fig:2}, we distinguish two groups of contributions to the recirculated power: those that are on only when the reactor is producing power, and those that are also on during the short maintenance periods.
\begin{enumerate}[i]
\item $P_{\text{e,recirc,op}}$ is needed only during a period of operation. A large part of this is the electric power needed to drive the toroidal plasma current 
\begin{equation}\label{ThermalPower}
P_\text{e,CD} = \eta_{\text{CD}} P_{\text{CD}},
\end{equation}
\noindent where $P_{\text{CD}}$ is the current drive power delivered to the plasma, and $\eta_{\text{CD}}$ is the wall-plug efficiency of the current drive system. The current drive power $P_{\text{e,CD}}$ importantly does not scale with the output power, as the reactor is designed to run a fixed values of current and magnetic field (to good approximation, cf. footnote \ref{firstapprox}). This group also includes the electric power needed to pump the coolant $P_{\text{e,pump}}$ and all other forms of power consumption -- mostly relatively small -- needed only during operation. Note that in the case of helium cooling the pumping power is considerable, typically estimated at 5--10$\%$ of $P_{\text{th}}$ \cite[p. 99]{PPCS}. Note that if the coolant is kept running during the short maintenance periods, the pumping power should be placed in group (ii), which increases the effective recirculated power fraction.

\item  $P_{\text{e,recirc,maint}}$  is kept on during short maintenance periods. This includes the power consumed by the cryoplant, which cools the superconducting magnets, the heat shield and the cryopumps. Group (ii) also encompasses all other power needed to maintain the plant, the balance of plant (BoP), including the active gas handling plant, the operation of the hot cell, the heating and air-conditioning of the plant buildings, the power demands of the computer systems, the power demands of the many auxiliary systems such as the vacuum system, the diagnostics, pellet injectors, etc.
\end{enumerate} 
\noindent This situation is represented by the two levels of recirculated power in Figure \ref{fig:2}. Using the maintenance parameter $\mu$ (defined above as the fraction of time during an available period that the plant is unproductive due to, for example, short maintenance), we define the  recirculated power averaged over periods of availability by 
\begin{equation}\label{RP}
\Braket{P_ {\text{e,recirc}}} =  (1 - \mu)P_{\text{e,recirc,op}} + {\mu P_{\text{e,recirc,maint}}},
\end{equation}
\noindent and finally the  \textit{effective} recirculated power fraction $f_{\text{recirc,eff}}$ as the ratio of the recirculated electric power and the nominal gross electric power, averaged over a period of time between major shutdowns:
\begin{equation}\label{RPFE}
f_ {\text{recirc,eff}}=  \Braket{P_{\text{e,recirc}}}/P_{\text{e,gross,nom}}.
\end{equation}
%\noindent Note that if the plant maintenance takes up only a small fraction of the time,  $f_{\text{recirc,eff}}$ is almost equal to the recirculated power fraction during operation. The maintenance periods reduce $f_{\text{recirc,eff}}$, while reducing $\chi$ at the same time. This means that the effect of maintenance on the plant efficiency is limited, but it does of course affect the plant economics.\\
We will use the effective recirculated power fraction in the following evaluation of the plant efficiency.

\subsection{Estimation of the expectation value of the effective recirculated power fraction}
Estimations for these power costs are given by the power plant conceptual studies (PPCS) \cite{PPCS} as well as in various later studies \cite{Pamela, Stork, Kovari}. We compared the numbers for early PPCS models A and B and the model power plants discussed in references \cite{Pamela, Stork} and calculated  $f_{\text{recirc,eff}}$ using as much as possible the same set of assumptions for all cases. These assumptions concern the current drive (wall plug) efficiency (PPCS and Pamela et al \cite{Pamela}  take 0.6, Stork \cite{Stork} takes 0.33; we have chosen 0.4 as an intermediary value); 
%XXX Discussion point - we could also take 0.33
and the BoP (we included consistent estimations for the cryoplant, the pumping power, and auxiliary systems in all cases).\footnote{To estimate the power demand of the cryoplant, we have scaled the power demand with the to-be-cooled area (major radius of the reactor squared), starting by identifying the cryoplant costs of ITER (30 MW) with that of PPCS model D, which have comparable major radii. This is consistent with Pamela et al \cite{Pamela}. We also note that PPCS give a similar number, but choose to not include it in the analysis because of the uncertainty.} This comparison is summarised in Table \ref{table:1}. We conclude that for a harmonised set of assumptions, all three studies come to a comparable value of the recirculated power fraction during operation.

 In conclusion, drawing on various sources and estimations, we find that for the continuous operation tokamak concept,  $f_{\text{recirc,eff}}$ should be expected to be in the range 0.4--0.6 for early generations of fusion power plants. This estimation is reflected by the shaded area in Figure \ref{fig:1}.

\begin{table}[h]
\centering
\includegraphics[width=0.75\textwidth]{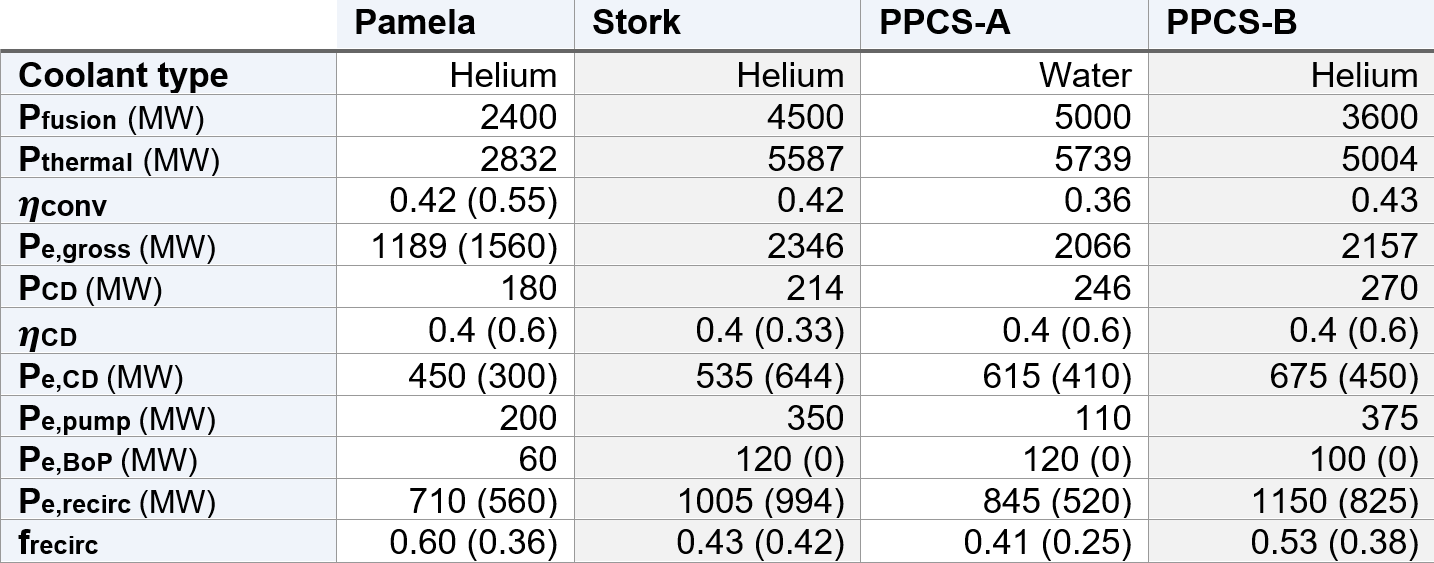}
\caption{\small{Comparison of estimates of the recirculated power in references \cite{Stork, Pamela, PPCS}, where we have applied consistent assumptions (in brackets the original numbers), showing a recirculated power fraction  between 0.4 and 0.6.}} \label{table:1}
\end{table}
\begin{figure}[h]
\centering
        \includegraphics[width=0.6\textwidth]{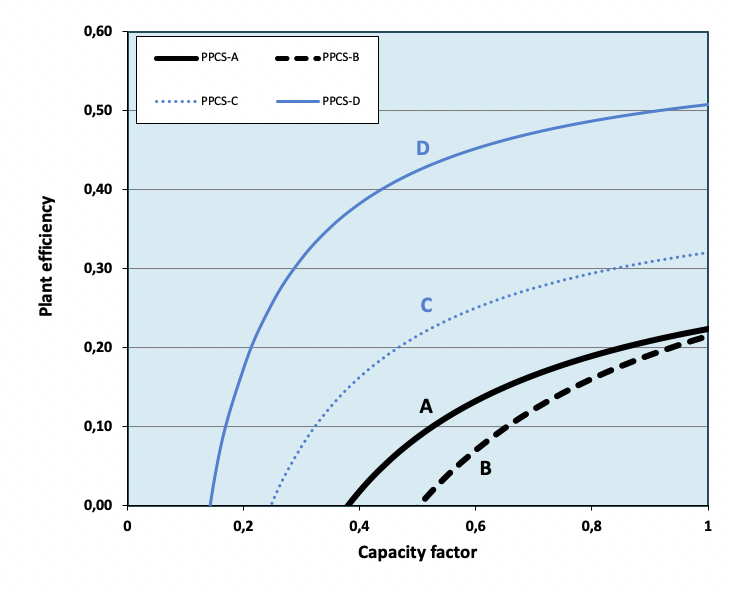}
\caption{\small{The plant efficiency of the models A to D of the PPCS, recalculated as a function of the capacity factor. Note that the values of $\eta_{\text{plant}}$  at nominal capacity ($\chi=1$) are lower than those in PPCS, which normalises the net output power to the thermal power generated in the reactor core only, where we have included all thermal power, i.e. including the nuclear power generated in the blanket. For the recirculated power calculation we used the same systematic as for the comparison in Table \ref{table:1}, resulting in recirculated power fractions of 0.34; 0.5; 0.25 and 0.14 for the models A--D, respectively. Other relevant parameters are: current drive efficiency $\eta_{\text{CD}}$ = 0.4; maintenance parameter $\mu$= 0.1 (i.e., the capacity factor cannot exceed 0.9)}.}
     \label{fig:4}
\end{figure}

\section{Plant efficiency} \label{PE}
Thermodynamically, the  plant efficiency $\eta_{\text{plant}}$ is the ratio between the net electric power that is delivered to the grid and the total thermal power that is produced before it is converted into electric power, i.e.,  
\begin{equation}\label{PEconcept}
\eta_{\text{plant}}=\frac{ P_{\text{e,net}} }{ P_{\text{th}}}=\frac{ P_{\text{e,gross}} - P_{\text{e,recirc}} }{ P_{\text{th}}},
\end{equation}

\noindent  This definition is similar to that of PPCS, except that in PPCS $P_{\text{therm}}$ only includes $P_{\text{fusion}}$ and excludes $P_{\text{breed}}$. It is a matter of definition, but we choose to normalise the net output to the total thermal power -- that is, representative of thermodynamic efficiency -- since the power produced in the blanket is (i) appreciable and (ii) an integral part of the plant economy in that it also consumes fuel (lithium and beryllium). 

With the definitions of the recirculated power fraction and the capacity factor given above, Eq. \eqref{PEconcept} translates to Eq. \eqref{efficiencyreduction}. As noted before, in the analysis of the fusion power plant we must use the \textit{effective} recirculated power fraction, averaged over a period of time between major shutdowns. %It is here that we see the combined effect of recirculated power and capacity as $\Braket{P_{\text{e,recirc}}}=\chi \times P_{\text{e,recirc}}$.

Figure \ref{fig:4}  shows how the inclusion of the capacity factor in the evaluation of the plant efficiency works out for the power plant models in PPCS. We see that the early models A and B are expected to have plant efficiencies well below 20$\%$ for realistic values of the capacity factor. The advanced models C and especially D are characterised by lower recirculated power and are therefore less sensitive to the capacity factor, as long as it remains above 50$\%$.

\FloatBarrier
\section{Implication of plant efficiency on the economics of fusion}
The previous sections established how recirculated power and capacity factor affect the efficiency of fusion power plants. The efficiency ultimately determines the amount of electricity that can be delivered to the grid and consequently the economic performance of fusion power plants. One common indicator for  the economic performance is the levelised cost of electricity (LCOE), which is defined as the average revenue per unit of energy output expressed in $\$/MWh$ and represents the lifetime average cost of energy for a specific power plant project. In a simplified form, the LCOE is formulated as the cost to build and operate a power plant over its lifetime divided by the total power output of the plant over that lifetime. That is,
\begin{equation}
\text{LCOE}=\frac{\text{sum of costs over lifetime}}{\text{sum of net electricity  produced over lifetime}}.
\end{equation}
This formula combines power plant ownership costs (capital and operating) and thermal performance (output and efficiency). We note that in previous sections we have restricted the analysis of the plant efficiency to the periods in which the plant is available. For an accurate LCOE the Availability ($\alpha$) must be factored in as well.  

Figure \ref{fig:4} shows that for the early PPCS models A and B, efficiencies below 20$\%$ can be expected. For fusion power plants, assuming the lifetime costs are to good approximation independent of the power produced, the LCOE will -- in first approximation -- be inversely proportional to $\alpha * \eta_{\text{plant}}$.   Again we stress that  for a given realistic recirculated power fraction, $\eta_{\text{plant}}$ depends much more sensitively on the capacity factor than might have been realised. 

Another important factor is the  relatively high fraction of the cost of capital of fusion power plants, due to the large initial investment. Fusion plants have relatively low variable cost, which is a good feature generally, but it magnifies the impact of efficiency  and availability on the LCOE. 

Measures such as the LCOE are a useful tool to analyse the competitiveness of fusion power once it is a well established component of the energy mix, i.e., when the deployment has reached the second or third generation of fusion plants, the technological -- and thereby the economical -- risks have been greatly reduced and the industrial capacity to produce power plants is well developed. 

However, as outlined in the introduction, the analysis in this paper is especially pertinent to the first generation of power plants. The low effective plant efficiency of early model fusion power plants predicted in this paper, coupled to the expected high investment costs as it is envisioned today \cite{Roadmap}, will create a competitive disadvantage. This also underscored the concern about the `valley of death' in the development of fusion power, which is the period while there is not yet a return on investment, but when a large investment is nevertheless needed for the construction of early generations of fusion reactors \cite{ValleyOfDeath}.  

\section{Discussion}
This paper draws attention to the fact that a fusion reactor, and certainly the first generation of fusion reactors, must be expected to, on average, run at a power below the nominal value. This has a number of consequences. First, while it was already observed in \cite{PPCS} and other literature that the recirculated power in a steady state tokamak reactor is undesirably high, it is the combination with a low capacity factor that will lead to a strong reduction of the power plant efficiency, jeopardising its economic viability. In practice, this means that it is imperative to take all thinkable measures to cap the recirculated power fraction. The continuously operating tokamak might not be economically viable. As a consequence, the pulsed tokamak, possibly with a modest amount of current drive to stretch the pulse, or -- in time -- the stellarator should be considered as the baseline scenario for at least the early commercial power plants.  

A second consequence is that the economics of the reactor are affected in various ways. The most natural way of expressing this is in the cost of electricity, which due to the recirculated power fraction shows a much stronger dependence on the capacity factor than the trivial reciprocal proportionality. Especially when both the capacity factor and the recirculated power approach 50$\%$, the cost of electricity becomes extremely sensitive to these parameters. This sensitivity is in itself highly unwanted, if a business case has to be made.
  
In a similar vein, a reduction of the plant efficiency impacts the energy payback time (EPBT), i.e., the time a power plant must operate before it returns the energy invested in constructing it with a cradle-to-grave system boundary \cite{Schleisner, Nicholas}, and thereby the attractiveness of fusion as a sustainable energy source.

Finally, the capacity factor affects the fuel cycle, in particular the tritium over-breeding. As the reactor requires the same start-up tritium inventory independent of the capacity factor it will achieve, the capacity factor directly affects the time the reactor has to run in order to breed enough tritium to start the next reactor \cite{Coleman}. 

Lower plant efficiency will drive up the levelised cost of electricity, which becomes very sensitive to the realised capacity factor. The uncertainty of the achievable capacity factor, and thereby the economic viability of the plant, could be expected to discourage private investors to participate in owning and operating fusion power plants. Also in the public perception, a large, complex and costly power plant which turns out to perform poorly, would not contribute to the support for fusion as an energy source. This would negatively affect government support for the deployment of fusion power. This analysis therefore stresses the importance of the development of plant designs with both  low recirculated power fraction and high capacity factor. This would, paradoxically, be most essential for the first generation of fusion power plants, precisely at the moment when these requirements are hardest to meet. But a failed first generation must be avoided if fusion is to see a second generation.

%An additional remark concerns the fact that PPCS is not fully consistent on on the blanket gain factor. The tritium-breeding reaction n(Li,He)T adds 4.8 MeV per reaction. Depending on the blanket configuration, the breeding ratio and neutron multiplication scheme, $P_{\text{breed}}$ is typically 10--40$\%$ of $P_{\text{DT}}$. PPCS Table 1 \cite[p. iv]{PPCS2} gives 18$\%$ for model A, 39$\%$ for model B, and 19$\%$ for models C and D. As a zero-order approximation one may take the value that would be achieved if each fusion-generated neutron would react with lithium to produce one new tritium, i.e., without the use of neutron multiplication and with a breeding ratio of 1, resulting in a blanket gain of (17.6 + 4.8)/17.6 = 1.27.  The second version of Table 1 in the PPCS report \cite[p. 8]{PPCS2} gives values of the `blanket gain', which are the values quoted. However, Table 2 (found in \cite[p. 13]{PPCS2}, reproduced here as Table \ref{table:1}) gives values of the Blanket Power as well as the Fusion Power and Divertor Power, from which the blanket gain can also be derived, resulting in blanket gains of 1.15; 1.39; 1.17; 1.10, for models A to D, respectively. For models B and C these estimates are fully consistent, but for models A and D the blanket gain from Table 2 (in PPCS report)  comes out smaller.

\section{Conclusion}
A commercial tokamak-based fusion power plant will on average operate below the nominal power, even when the long shutdown periods for large maintenance are left outside the averaging, resulting in a capacity factor lower than one. Conventional and other renewable energy sources have low recirculated power, so that the capacity factor is of little concern. In this paper, we have shown how for fusion the combination of high recirculated power and a low capacity factor drives down the plant efficiency significantly. We argue that this will  be a major concern for fusion power plants based on a steady state tokamak concept. Low recirculated power and high capacity factor will be crucial for the success of the first generation of fusion power plants.

\section{Acknowledgement}
The authors thank Guido Lange for valuable discussions and suggestions for visualisation. We also thank two anonymous referees for clarification of concepts  and suggesting valuable literature.

 \footnotesize{ }

\end{document}